\documentclass[showpacs,aps]{revtex4}
\usepackage{amsmath}
\usepackage{times}
\usepackage{epsfig}
\usepackage{amssymb}
\usepackage{graphicx}
\usepackage{pstricks}
\usepackage{psfrag}

\begin{document}
\today
\title{
Current Fluctuations of the One Dimensional Symmetric Simple Exclusion Process
with a Step Initial Condition 
}
\author{Bernard Derrida and Antoine Gerschenfeld}
\affiliation{ Laboratoire de Physique Statistique, Ecole Normale
Sup\'erieure,
24 rue Lhomond, 75231 Paris Cedex 05 - France}
\begin{abstract}
For the symmetric simple exclusion process on an infinite line, we
calculate exactly the fluctuations of the integrated current $Q_t$ during time
$t$ through the origin when, in the initial condition, the sites are occupied
with density $\rho_a$ on the negative axis and with density $\rho_b$ on the
positive axis.
All the cumulants of $Q_t$ grow like $\sqrt{t}$. In the range where $Q_t \sim
\sqrt{t}$, the decay $\exp [-Q_t^3/t]$ of the distribution of $Q_t$ is
non-Gaussian.
Our results are obtained using the Bethe ansatz and several identities
recently derived by Tracy and Widom for exclusion processes on the infinite
line.
\end{abstract}

\maketitle

\date{\today}

key words: current fluctuations - Bethe ansatz - Symmetric exclusion

\section{Introduction}
\label{intro}
Understanding how currents of particles or of heat fluctuate through non
equilibrium systems has motivated lots of efforts over the last two decades.
A number of general symmetries of the distribution of these fluctuations,
based on the microreversibility of the dynamics, have been discovered
\cite{ECM,GC,Jarz,Crooks}.
Beyond these symmetries, the distribution of the current fluctuations has
been calculated in several cases \cite{PS2,BD,HRS1,HRS2,GKP,HG}.
For diffusive systems, a general theory, the macroscopic fluctuation theory,
has been developed \cite{BDGJL2} which, under some conditions
\cite{BDGJL5,BDGJL6,BD2005}, allows one to calculate the whole distribution of
these current fluctuations for systems maintained in a non-equilibrium steady
state by contact with two reservoirs (of particles at different chemical
potentials or of heat at different temperatures) \cite{BD,HG,derrida2007}.
For driven systems, which are not diffusive, the fluctuations belong to
the universality class of the Kardar Parisi Zhang equation \cite{KPZ} and can
be related to the fluctuations in growth models and to the theory of random
matrices \cite{J,PS,PS2,sasamoto,TW1,TW2,RS}.

For diffusive systems as well as for driven systems, it has been very helpful
to analyse simple models, in particular exclusion processes
\cite{HS,Liggett2,KL}.
In exclusion processes, one considers particles hopping on a lattice with a
hard core interaction which prevents two particles from being on the same
lattice site.
In one dimension, exclusion processes can be solved by the Bethe ansatz
\cite{gwaspohn,DL,OG2,Degier1,Degier2,TW1,TW2,TW3,PM,ADLW} and a number of
exact results have been obtained on the fluctuation of the current or the
distribution of particles for a given initial condition
\cite{schutz2,Priezzhev,PP,DP}.

In the present work we consider the symmetric simple exclusion
(SSEP) on an infinite one dimensional lattice.
By definition of the model each lattice site is either empty or occupied by a
single particle.
The dynamics is stochastic: each particle hops to each of its neighboring
sites at rate 1 if the move is not forbidden by the exclusion rule (each site
is occupied by at most one particle).
At time $t=0$ each site at the left of the origin ($x \leq 0$) is occupied
with a probability $\rho_a$ and each site at the right of the origin ($x>0$)
is occupied with a probability $\rho_b$ (see figure 1).
We also assume that the measure at $t=0$ is Bernoulli, meaning that there are
no correlation between the occupation numbers of the different sites in the
initial condition.
We call $Q_t$ the total flux of particles between site $0$ and site $1$ during
the time interval $t$. Our goal in the present paper is to determine, for
large times $t$, the probability distribution of $Q_t$.

\begin{figure}[ht]
\centerline{\includegraphics[width=8cm]{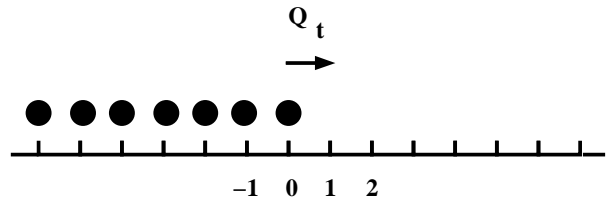}}
\caption{The initial condition when $\rho_a=1$ and $\rho_b=0$}
\label{initial}
\end{figure}

The same initial condition has already been considered for the asymmetric
simple exclusion (ASEP) \cite{schutz2,PS2,sasamoto,IS,TW1,TW2,AKR,TW3}, in
particular in connection to random matrix theory.

\bigskip

Our main result is that, for large $t$, the generating function of this total
flux is given by
\begin{equation}
\left\langle e^{\lambda Q_t} \right\rangle \sim  e^{\sqrt{t} F(\omega) }
\label{res1}
\end{equation}
where the function $F(\omega)$ is defined by
\begin{equation}
F(\omega) = {1 \over \sqrt{\pi}}\sum_{n \geq1} {( -)^{n+1} \over{n}^{3/2}} \
\omega^n \equiv {1 \over {\pi}} \int_{ \infty}^\infty dk \log \left[ 1 +
\omega e^{-k^2} \right]
\label{Fdef}
\end{equation}
and $\omega$ is a function of $\rho_a, \rho_b, \lambda$
\begin{equation}
\omega= \rho_a( e^\lambda -1) + \rho_b(e^{-\lambda}- 1) + \rho_a \rho_b ( 
e^\lambda -1) ( e^{-\lambda} -1) \,.
\label{omega-def}
\end{equation}
{\bf Remark:}
It has already been noticed \cite{DDR} that in the steady state of the SSEP on
a large but finite chain, the generating function of the current was also a
function of $\rho_a, \rho_b, \lambda$ through the same single parametrer,
$\omega$. This is a general property of the symmetric exclusion process that
we shall establish in section 2 for a general graph.
\ \\ \ \\
{\bf Remark:} From (\ref{res1},\ref{Fdef},\ref{omega-def}) one can  obtain
all the cumulants of $Q_t$, in the large $t$ limit. They all grow like
$\sqrt {t}$ and the prefactor in the cumulant $\langle Q_t^n \rangle_c$
is a polynomial of degree $n$ in $\rho_a$ and $\rho_b$.
For example, for $\rho_a=\rho$ and $\rho_b=0$, one gets
\begin{eqnarray*}
   \lim_{t\to \infty} t^{-1/2} \langle Q_t \rangle_c &=&{1 \over \sqrt{\pi}} \ 
  \rho\,,\\
  \lim_{t\to \infty} t^{-1/2} \langle Q_t^2 \rangle_c &=& {1 \over
  \sqrt{\pi}} \left(\rho- {\rho^2 \over \sqrt{2}}\right)\,,\\
   \lim_{t\to \infty} t^{-1/2} \langle Q_t^3 \rangle_c &=& {1 \over
  \sqrt{\pi}} \left(\rho-  {3 \over \sqrt{2}} \ \rho^2  + {2 \over \sqrt{3}} \ 
  \rho^3 \right)\,,\\
\end{eqnarray*}

 while for $\rho_a= \rho_b  =\rho$ one gets for the even cumulants
(all the odd ones  vanish by symmetry)
\begin{eqnarray*}
  \lim_{t\to \infty} t^{-1/2} \langle Q_t^2 \rangle_c &=& {2 \over \sqrt{\pi}}  
  \ \rho(1- \rho)\,,\\
  \lim_{t\to \infty} t^{-1/2} \langle Q_t^4 \rangle_c &=& {2 \over
  \sqrt{\pi}}  \ \rho(1- \rho) - {6 \sqrt{2} \over \sqrt{\pi}}  \  \rho^2(1- 
  \rho)^2\,.
\end{eqnarray*}
This can be compared to the variance of the position $X_t$ of a tagged
particle \cite{richards,Alexander,Arratia,Saada,VKK,Kumar}, which also grows
like $\sqrt{t}$, by arguing that the typical distance between particles is
$\rho^{-1}$ so that $\langle Q_t^2 \rangle \simeq \rho^2 \langle X_t^2
\rangle$.

From the knowledge of the generating function (\ref{res1}), one can obtain the
distribution of $Q_t$. This distribution takes, in the long time limit, a
scaling form
\begin{equation}
{\rm Pro} \left( {Q_t \over \sqrt{t}} = q \right) \sim \exp[\sqrt{t} G(q) ]\,,
\end{equation}
where $G(q)$ can be related to $F(\omega)$ by a Lagrange transform. This
allows one to obtain the asymptotics of $G(q)$. One can see from the integral
representation (\ref{Fdef}) of $F(\omega)$ that, for large positive $\omega$,

$$F(\omega) \simeq {4 \over 3 {\pi}} \  [ \log \omega ]^{3/2} +  {\pi \over 6 
}  \ [\log \omega ]^{-1/2}  + ..$$

and this gives for large positive $q$

\begin{equation}
G(q) \simeq -{\pi^2 \over 12} \  q^3 + q \log(\rho_a(1-\rho_b) )  +..
\label{q3}
\end{equation}

This paper is organized as follows: in section II we prove that the generating
function $\left\langle e^{\lambda Q_t} \right\rangle$ of $Q_t$ is a function
of the densities $\rho_a, \rho_b$ and of $\lambda$ through the single
parameter $\omega$ defined in (\ref{omega-def}).
In section III we obtain an exact expression of the generating function
of $Q_t$ from which  we derive the large $t$ asymptotics (\ref{res1},\ref{Fdef}). 

\begin{figure}[ht]
\centerline{\input{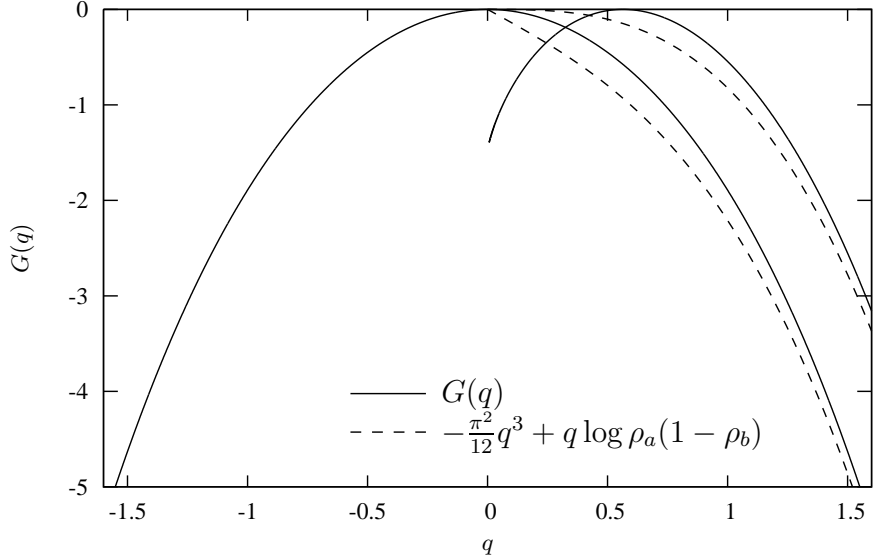}}
\caption{The large deviation function $G(q)$ and its asymptotics for large $q$ (dashed lines) in the cases $\rho_a = \rho_b= 1/2$ (left curves) and $\rho_a = 1$, $\rho_b=0$ (right curves).}
\label{fig-largedev}
\end{figure}

\section{Evolution of the correlation functions and the $\omega$
dependence}
\label{sec:1}
Let $\tau_x$ be a binary variable which indicates whether site $x$ is occupied
($\tau_x=1$) or empty ($\tau_x=0$).
In this section we are going to argue that all the $n$-point correlation
functions of these $\tau_x$'s are polynomials of degree $n$ in $\rho_a$ and
$\rho_b$.
This will allow us to show that the $n$-th moment $\langle Q_t^n \rangle$ is
also a polynomial of degree $n$ in $\rho_a$ and $\rho_b$.
Then using the particle-hole symmetry, we will show that the generating
function $\left\langle e^{\lambda Q_t} \right\rangle$ depends on
$\rho_a,\rho_b$ and $\lambda$ through the single parameter $\omega$ defined in
(\ref{omega-def}).

One can write the exact evolution of the expectation $\langle \tau_x \rangle$
where $\langle.\rangle$ denotes an average over the stochastic evolution and
the initial conditions.
On the infinite line it takes the form
\begin{equation}
{d \over dt} \langle \tau_x \rangle = \langle \tau_{x+1} \rangle
+ \langle \tau_{x-1} \rangle - 2 \langle \tau_x \rangle \,.
\label{one_point}
\end{equation}
As at time $t=0$, one has $\langle \tau_x \rangle= \rho_a$ for $x \leq 0$ and
$\langle \tau_x \rangle= \rho_b$ for $x > 0$, it is clear that at any later
time the solution of (\ref{one_point}) is linear in $\rho_a$ and $\rho_b$.

Similarly for any $x < y$ the evolution of the 2-point correlation function is
given for $y> x+1$ by
\begin{equation}
{d \over dt} \langle \tau_x \tau_y \rangle = \langle \tau_{x+1} \tau_y \rangle
+\langle \tau_{x-1} \tau_y \rangle +\langle \tau_{x} \tau_{y+1} \rangle
+\langle \tau_{x} \tau_{y-1} \rangle - 4 \langle \tau_x \tau_y\rangle
\label{two_point_1}
\end{equation}
and  for $y=x+1$ by
\begin{equation}
{d \over dt} \langle \tau_x \tau_{x+1} \rangle = \langle \tau_{x-1} \tau_{x+1}
\rangle +\langle \tau_{x} \tau_{x+2} \rangle - 2 \langle \tau_x
\tau_{x+1}\rangle\,.
\label{two_point_2}
\end{equation}
These evolution equations do not involve higher correlation functions and as
at $t=0$ the only possible values of $\langle\tau_x \tau_y \rangle $ are
$\rho_a^2, \rho_a \rho_b$ or $ \rho_b^2$, one can easily see that at any later
time $ \langle \tau_x \tau_y \rangle $ remains a quadratic function of
$\rho_a$ and $\rho_b$.

One can generalize this property of the SSEP to all equal-time or unequal-time
correlation functions on a general graph. Let us consider the symmetric
exclusion process on such a graph : the vertices are either empty or occupied
by a single particle and each particle can hop at rate one to every empty site
to which it is directly connected by the graph.
Therefore if $X\equiv \{ x_1, .. x_n\}$ is a set of $n$ different sites on
the graph and if $\langle \tau_X \rangle$ denotes their $n$-point correlation
function
$$\langle \tau_X \rangle = \langle \tau_{x_1} ... \tau_{x_n} \rangle \,,$$
it is easy to see that the evolution of $\langle \tau_X \rangle$ is given by
\begin{equation}
{ d \langle \tau_X \rangle \over dt} = \sum_{y \not\in X} \sum_{k=1}^n
\chi(y,x_k)[\langle \tau_y\tau_{X \setminus \{x_k\}} \rangle - \langle \tau_X
\rangle]
\label{evol2}
\end{equation}
 where $\chi(x,y)=1$ or $0$ indicates whether the edge $(x,y)$ is present or
not on the graph.
If initially all the $\tau_x$'s are uncorrelated, some sites being occupied
with probability $\rho_a$ and all the other sites occupied with probability
$\rho_b$, all the equal-time $n$-point correlation functions are obviously
polynomials of degree exactly $n$ in $\rho_a$ and $\rho_b$ at time $t=0$.
This property is preserved by the evolution (\ref{evol2}).
A similar reasoning allows one to show that all the unequal time $n$-point
correlation functions are also polynomials in $\rho_a$ and $\rho_b$ of degree
at most $n$.

On this general graph let us call $A$ the subset of sites occupied at time
$t=0$ with probability $\rho_a$ (all the other sites being occupied with
probability $\rho_b$). Obviously one has $\langle \tau_x \rangle = \rho_a$ if
$ x \in A$ and $\langle \tau_x \rangle = \rho_b$ if $ x \not\in A$. If $Q_t$
is the total flux out of the subset $A$ during time $t$, one has
\begin{equation}
Q_t = \sum_{x \in A} \tau_x(0)- \tau_x(t) \,.
\label{Qt-def}
\end{equation}
It is clear from (\ref{Qt-def}) that $\langle Q_t^n \rangle$ can be expressed
in terms of unequal-time $n$-point correlation functions and is therefore a
polynomial of degree $n$ in $\rho_a$ and $\rho_b$ (if one or more sites are
repeated more than once in the correlation function, this may give a
polynomial of lower degree as $\tau_x^k = \tau_x$ for all $k \geq 1$).

\ \\ \ \\ \

To show that $\langle e^{\lambda Q_t}\rangle$ depends only on the reduced
parameter $\omega$, let us consider a finite lattice of $N$ sites, where
initially all the $N_A$ sites of the subset $A$ are occupied with density
$\rho_a$, and all the remaining $N-N_A$ sites are occupied with density
$\rho_b$.
One can write the generating function $\left\langle e^{\lambda Q_t}
\right\rangle$ as
\begin{equation}
\left\langle e^{\lambda Q_t} \right\rangle = \sum_{p=0}^{N_A}
\sum_{q=0}^{N-N_A} \rho_a^p (1-\rho_a)^{N_A -p} \ \ \rho_b^q (1-\rho_b)^{N-N_A
-q} \ \ e^{-q \lambda} R_{p,q} ( e^\lambda) \,,
\label{gen1}
\end{equation}
where $R_{p,q}(z)$ is a polynomial of degree $p+q$ in $z$ which depends on
time and on the choice of the subset $A$.
The reason for the factor $ e^{-q \lambda} R_{p,q} ( e^\lambda)$ is
that this term corresponds to situations where initially there are $p$
particles in the subgraph $A$ and $q$ particles in the rest of the graph
so that $Q_t$ can only take the values $Q_t=-q,-q+1,.. p$.
If one expands this expression in powers of $\rho_a$ and $\rho_b$, one gets
\begin{equation}
\left\langle e^{\lambda Q_t} \right\rangle =\sum_{p=0}^{N_A}\sum_{q=0}^{N-N_A}
 \rho_a^p \ \rho_b^q \ e^{-q \lambda} S_{p,q} ( e^\lambda)\,,
\end{equation}
where  $S_{p,q}(z)$ is also a polynomial of degree $p+q$.
For $\langle Q_t^n \rangle$ to be a polynomial of degree $n$ in
$\rho_a $ and $ \rho_b$ one needs that for $\lambda$ small
$$S_{p,q} ( e^\lambda) = O(\lambda^{p+q})\,,$$
so that the polynomial $S_{p,q}$ has the form
$$S_{p,q} ( e^\lambda) = s_{p,q} (e^\lambda-1)^{p+q} \,,$$
where $s_{p,q}$ is a number which depends on time, on the graph and on the
subgraph $A$ but does not depend on $\lambda$.
This implies that (\ref{gen1}) can be rewritten as
\begin{equation}
\left\langle e^{\lambda Q_t} \right\rangle = \sum_{p=0}^{N_A}
\sum_{q=0}^{N-N_A} s_{p,q} \ \ [\rho_a( e^\lambda-1)]^p \ \ [\rho_b(
e^{-\lambda}-1)]^q \equiv G\big( \rho_a( e^\lambda-1), \rho_b(
e^{-\lambda}-1)\big) \,.
\label{G-def}
\end{equation}
This shows clearly that $\left\langle e^{\lambda Q_t} \right\rangle$ is
already a function of only two reduced variables: $\rho_a( e^\lambda-1)$ and
$\rho_b( e^{-\lambda}-1)$.

Let us now use the particle-hole symmetry. In the SSEP, holes have exactly the
same dynamics as particles. Therefore the generating function of the flux
$Q_t$ is left invariant by the particle-hole symmetry $(\rho_a,\rho_b,\lambda)
\to (1-\rho_a,1-\rho_b,-\lambda)$.
In terms of the function $G$ of two variables defined in (\ref{G-def}), this
means that, for any $\rho_a, \rho_b,\lambda$
\begin{equation}
G\big( \rho_a( e^\lambda-1), \rho_b( e^{-\lambda}-1)\big) =
G\big( (1-\rho_a)( e^{-\lambda}-1), (1-\rho_b)( e^{\lambda}-1)\big)\,, 
\label{G-sym}
\end{equation}
and so, if $\alpha= \rho_a( e^\lambda-1)$ and $\beta= \rho_b( e^{-\lambda}-1)$
, one has
\begin{equation}
G\big( \alpha, \ \beta\big) = G\big( e^{-\lambda}-1 + \alpha e^{-\lambda},
e^{\lambda}-1 + \beta e^{\lambda}\big)\,.
\label{G-sym1}
\end{equation}
As (\ref{G-sym1}) is valid for any $\lambda$, one can choose $e^{-\lambda}= 1+
\beta$, which leads to
\begin{equation}
G\big( \alpha, \ \beta\big) = G\big( \alpha + \beta + \alpha \beta, 0\big)\,.
\end{equation}
This completes the proof that $G(\alpha,\beta)$ and therefore $\left\langle
e^{\lambda Q_t} \right\rangle$ are functions of the single variable \\
$\alpha + \beta + \alpha \beta= \rho_a( e^\lambda-1) +\rho_b( e^{-\lambda}-1)+
\rho_a( e^\lambda-1) \rho_b( e^{-\lambda}-1) \equiv \omega$.

\ \\
{\bf Remark:} 
This $\omega$ dependence of $\left\langle e^{\lambda Q_t} \right\rangle$ was
already noticed for the SSEP on a finite lattice of length $L$ connected at
its extremities to two reservoirs of particles in \cite{DDR}, where it
was, however, only obtained in the large $t$ and $L$ limit.
A consequence of the above discussion is that the $\omega$ dependence remains
valid for any system size $L$, at any time $t$, in more complicated
geometries (in particular in higher dimensions), and for more
complicated connections to the two reservoirs at densities $\rho_a$ and
$\rho_b$.
To illustrate this claim, let us consider as in \cite{DDR} the SSEP on a one
dimensional lattice of $L$ sites, with particles injected on site $1$ and
$L$ at rates $\alpha$ and $\delta$ and removed from these two sites at rates
$\gamma$ and $\beta$.
It is well known that these rates correspond to site $1$ being connected to a
reservoir at density $\rho_a= {\alpha \over \alpha + \gamma}$ and site $L$ to
a reservoir at density $\rho_b= {\delta \over \beta + \delta}$.
Instead of these input rates, one could think of site $1$ being connected to a
large number $N_1$ of sites (which mimic the left reservoir), which are all
initially at density $\rho_a$, with an exchange rate $ (\alpha+\gamma)/N_1$
between site $1$ and each of these $N_1$ sites of the reservoir.
Similarly one can replace the rates $\beta$ and $\delta$ by a large reservoir
of $N_L$ sites, initially at density $\rho_b= {\delta \over \beta + \delta}$,
with an exchange rate $(\beta+ \delta)/N_L$ with site $L$.
If inially all sites $i \leq i_0$ are at density $\rho_a$ and all sites $i
\geq i_0+1$ are at density $\rho_b$, we are in a situation where all the sites
of the graph composed by the one-dimensional lattice and the reservoirs are
initially occupied with probability either $\rho_a$ or $\rho_b$.
In this geometry the flux $Q_t$ is then simply the total flux of particles
between site $i_0$ and site $i_0+1$, and therefore, for this flux, we know
from (\ref{G-sym}) that $ \left\langle e^{\lambda Q_t} \right\rangle$ depends
on $\rho_a,\rho_b$ and $\lambda$ through the single parameter $\omega$.
\ \\ \ \\ 

Because the generating function $ \left\langle e^{\lambda Q_t} \right\rangle$
depends only on the single parameter $\omega$ defined in (\ref{omega-def}), it
can be written as
\begin{equation} 
\left\langle e^{\lambda Q_t} \right\rangle = \sum_{n \geq 0} \ s_n(t) \
\omega^n \ .
\label{expansion}
\end{equation}
Therefore, to determine the coefficients $s_n(t)$, one can limit the
discussion to the particular case $\rho_b=0$ where the analysis is easier.
Then
\begin{equation}
\left\langle e^{\lambda Q_t} \right\rangle = \sum_{n \geq 0} s_n(t)\  \rho_a^n\  (e^\lambda-1)^n \,.
\label{expansion-1}
\end{equation}
This can be viewed as an expansion of $\left\langle e^{\lambda Q_t}
\right\rangle$ in powers to $\rho_a$. We are now going to express the
coefficients $s_n(t)$ in terms of properties of exclusion process on an
arbitrary graph.
For simplicity we consider a finite graph.
Initially, only the sites of the subset $A$ are occupied, with probability
$\rho_a$ : very much like in (\ref{gen1}) one can write the generating
function $\left\langle e^{\lambda Q_t} \right\rangle$ as
\begin{equation}
\left\langle e^{\lambda Q_t} \right\rangle= \sum_{E \subseteq A} \rho_a^{|E|}
(1- \rho_a)^{N_A- |E|} \ \sum_{q=0}^{|E|} {\rm Pro}_t(E,q) \ e^{\lambda q}\,,
\label{expansion-2}
\end{equation}
where ${\rm Pro}_t(E,q) $ is the probability that $q$ particles have escaped
from $A$ during $t$, given that at $t=0$ only its subset $E$ was occupied by
particles.
Comparing (\ref{expansion-1}) and (\ref{expansion-2}), we see that the
coefficients $s_n(t)$ can be expressed in terms of the probabilities ${\rm
Pro}_t(E,q) $.
In fact, this relation takes a simple form, which can be easily understood by
taking simultaneously the limits $\rho_a \to 0$ and $\lambda \to \infty$, at
fixed $\rho_a e^\lambda$, in (\ref{expansion-1}) and (\ref{expansion-2}) : this leads to
\begin{equation}     
s_n(t)= \sum_{E \subseteq A,|E|=n} {\rm Pro}_t(E,n) \,,
\label{sn}     
\end{equation}          
where the sum is over all the subsets $E$ of $n$ sites and ${\rm Pro}_t(E,n)$
is the probability that, if initially all the $n$ sites of $E$ are occupied,
the rest of the graph being empty, all the $n$ particles have escaped from $A$
at time $t$.

\section{The Bethe ansatz}
\label{sec:2}
The evolution equations of the correlation functions, on the infinite line,
can be solved via the Bethe ansatz :
for a fixed initial configuration, i.e. $\{\tau_x(0)\}= \{\eta_x\}$, it allows
one to obtain the expression of all the correlations functions at any later
time. For example the solution of (\ref{one_point}) is
\begin{equation}
\langle \tau_y(t)\rangle = \sum_{x} P_t^{(1)}(y|x) \ \eta_{x}\,,
\label{one_point_sol}
\end{equation}
where
\begin{equation}
P_t^{(1)}(y|x)= \oint_{|z_1|=r} {dz_1 \over 2 i \pi z_1} \ z_1^{y-x} \ \exp
\left[\left( z_1+ {1 \over z_1}-2 \right )t\right]
\label{one_point_sol_1}
\end{equation}
where the integration contour is a counterclockwise circle of radius
$r$ (for convenience in what follows we choose $r \ll 1$).

Similarly one can show that the solution of
(\ref{two_point_1},\ref{two_point_2}) is, for $y_1<y_2$,
\begin{equation}
\langle \tau_{y_1}(t) \tau_{y_2}(t) \rangle = \sum_{x_1<x_2}
P_t^{(2)}(y_1,y_2|x_1,x_2) \ \eta_{x_1} \eta_{x_2}\,,
\label{two_point_sol}
\end{equation}
where
\begin{eqnarray}
P_t^{(2)}(y_1,y_2|x_1,x_2)=
\oint_{|z_1|=|z_2|=r} 
{dz_1  \over 2 i \pi z_1}
\ 
{dz_2  \over 2 i \pi z_2}
 \   \left[z_1^{y_1-x_1} z_2^{ y_2-x_2}
 -  \left({ z_1 z_2  + 1 - 2 z_2 \over  z_1 z_2 + 1 - 2 z_1} \right) 
   z_1^{ y_2-x_1} 
z_2^{y_1-x_2}
\right]
\nonumber \\
\ \ \times  \  \exp \left[\left( z_1 + {1 \over z_1} +z_2 + {1 \over z_2} -4
\right )t\right]\,.\;\;\;\;\;
\label{two_point_sol_1}
\end{eqnarray}
In fact (\ref{one_point_sol}) and (\ref{two_point_sol}) can be extended to
arbitrary correlation functions to give
\begin{equation}
\langle \tau_{y_1}(t) \tau_{y_2}(t) ..  \tau_{y_n}(t) \rangle = \sum_{x_1<x_2 < .. < x_n}
P_t^{(n)}(y_1,y_2,..,y_n|x_1,x_2, .. x_n)  \  \eta_{x_1} \eta_{x_2} .. \eta_{x_n}
\label{n_point_sol}
\end{equation}
where $P_t^{(n)}(y_1,y_2,..,y_n|x_1,x_2, .. x_n)$ has a Bethe ansatz
expression which generalizes (\ref{one_point_sol_1},\ref{two_point_sol_1})
(see Appendix \ref{Apx:1} and \cite{TW1}).

One can then use (\ref{sn}) to establish that the coefficients $s_n(t)$ are
given by
\begin{equation}
s_n(t) = 
 \sum_{x_1<x_2 < .. < x_n \leq 0 < y_1 < .. < y_n}
P_t^{(n)}(y_1,y_2,..,y_n|x_1,x_2, .. x_n) \,.
\label{sn_1}
\end{equation}
For example one gets from (\ref{one_point_sol_1}, \ref{two_point_sol_1}) for $n=1$ and $n=2$
\begin{equation}
s_1(t) = \oint_{|z_1|=r} {dz_1 \over  2 i \pi }
\  {1 \over (1- z_1)^2} 
 \ \exp \left[\left( z_1+ {1 \over z_1}-2
\right )t\right]
\label{s1}
\end{equation}
and 
\begin{eqnarray}
s_2(t)=
\oint_{|z_1| =|z_2| =r}
{dz_1  \over 2 i \pi z_1}
\
{dz_2  \over 2 i \pi z_2}
 \    \left[ 
{z_1^2 z_2^2 \over(1-z_1)(1-z_2)(1-z_1 z_2)^2}
 \right.\nonumber \;\;\;\;\;\;\;\;\;\;\;\;\;\;\;\;\;\;\;\;\;\;\;\;\;\;\;\;
 \;\;\;\;\;\;\;\;\;\;\;\;\;\;\;\;\;\;\;\;\;\;\;\;\;\;\;\;&& \\
 -\left.
{z_1^3 z_2 \over(1-z_1)^2(1-z_1 z_2)^2}
\left(  { z_1 z_2  + 1 - 2 z_2 \over  z_1 z_2 + 1 - 2 z_1} \right)
\right]\ \times  \  \exp \left[\left( z_1 + {1 \over z_1} +z_2 + {1 \over z_2} -4
\right )t\right]\,.\;\; &&
\label{s2}
\end{eqnarray}
By symmetrizing the above expression over $z_1$ and $z_2$ one can show that this can be rewritten as
\begin{eqnarray}
s_2(t)= {1 \over 2}
\oint_{|z_i| =r}
{dz_1  \over 2 i \pi }
\
{dz_2  \over 2 i \pi }
 \  \  \det \left({1 \over z_k z_l +1 - 2 z_l} \right)_{1\leq k,l \leq2}
 \   \exp \left[\left( z_1 + {1 \over z_1} +z_2 + {1 \over z_2} -4
\right )t\right]\,.
\label{s2-bis}
\end{eqnarray}

This expression can in fact be extended to arbitrary $n$, using an approach
which follows closely recent works by Tracy and Widom \cite{TW1,TW2}. The
derivation, which is detailed in Appendix \ref{Apx:1}, yields
\begin{eqnarray}
\langle e^{\lambda Q_t}\rangle = \sum_{n\geq0}\omega^n s_n(t)=
\sum_{n\geq0}{\omega^n \over n!}
\oint_{|z_k|= r} 
\left[\prod_{k=1}^n { dz_k  e^{t(z_k+1/z_k-2)}\over 2 i \pi }
\right]
\
 \  \  \det \left({1 \over z_k z_l +1 - 2 z_l} \right)_{1\leq k,l \leq n}
\label{sn-bis}
\end{eqnarray}
 \ \\   \ \\ \ \\
It is  known \cite{BSimon}  (see also eq.(7) of \cite{TW2}) that  for a Fredholm operator $K$ which transforms a function $f$ into a function $Kf$ by

$$Kf(z)= \int_{|z| = r} {dz \over 2 i \pi} K(z,z') f(z')\,,$$
one has
\begin{equation}
\det(I + \omega K) = 
\sum_{n\geq0}{\omega^n \over n!}
\oint_{|z_k| =r}  \ \left[\prod_{k=1}^n  {dz_k \over 2 i \pi}\right]
   \  \det \left(K(z_k,z_l) \right)_{1\leq k,l \leq n}\,.
\label{determinant}
\end{equation}
This implies that
\begin{equation}
\log [\det(I + \omega K) ]= {\rm tr}[ \log(I+ \omega K)] = -
\sum_{n\geq0}{(-\omega)^n \over n}
\oint_{|z_k| =r}
 \left[\prod_{k=1}^n  {dz_k \over 2 i \pi}\right]
K(z_1,z_2) .. K(z_n,z_1)\,.
\label{logdeterminant}
\end{equation}
Comparing (\ref{sn-bis}) and (\ref{determinant}), we see that by choosing 
\begin{equation}
K(z,z')= 
{\exp\left[\left( z + {1 \over z}
  -2\right)t\right]\over zz'+1-2z}\,,
\label{Kdef}
\end{equation}
one gets (\ref{logdeterminant}) that
\begin{eqnarray}
\log \langle e^{\lambda Q_t}\rangle = -
\sum_{n\geq 1}{(-\omega)^n \over n}
\oint_{|z_k|= r}
\left[\prod_{k=1}^n { 
dz_k  \over 2 i \pi }
\right]
\
 \  \   \prod_{k=1}^n \left(
{
 \exp \left[\left( z_k + {1 \over z_k} -2 \right )t\right]
 \over z_k z_{k+1} +1 - 2 z_{k+1}} \right)
\label{sn-ter}
\end{eqnarray}
(with the convention that $z_{n+1} \equiv z_1$ in the $n$-th term).

In Appendix B, we derive the large $t$ behavior of the integrals in
the r.h.s. of (\ref{sn-ter}), which leads to 
$$ \log\langle e^{\lambda Q_t}\rangle \sim
-\sum_{n\geq1}{(-\omega)^n\over n} \sqrt{t\over\pi n}  = \sqrt
{t} F(\omega)\,,$$
completing the derivation  of (\ref{res1},\ref{Fdef}).
\\
\\ \ \\ \ \\ {\bf Remark :}
from the expressions (\ref{one_point_sol}, \ref{one_point_sol_1},
\ref{two_point_sol},\ref{two_point_sol_1}, \ref{n_point_sol}, \ref{TWAnsatz}),
one can in principle calculate arbitrary correlation functions of the
occupation numbers $\tau_i$ at time $t$. For the
one-point and the connected two-point functions one gets when $\rho_a=1$
and $\rho_b=0$
$$\langle\tau_x\rangle = \oint {dz e^{t(z+1/z-2)}\over 2i\pi z}{ z^x
\over1-z}
\ \ \ \ \mbox{and} \ \ \ \ \langle\tau_x\tau_y\rangle_c = \oint {dz dz'
e^{t(z + 1/z + z' +
1/z' - 4)}\over 4\pi^2 z z'}{z^x z'^y \over zz'+1-2z'} \,.$$

Then, carrying out an asymptotic analysis similar to the one performed in
Appendix \ref{Apx:2} yields for $n=1$ and $n=2$

\begin{equation}
\langle \tau_{x_1}..\tau_{x_n}\rangle_c \simeq {t^{{1-n \over 2}}}
G_n\left({x_1\over\sqrt{t}},..,{x_n\over\sqrt{t}}\right) \,.
\label{scaling-n}
\end{equation}
with
\begin{equation}
G_1(X)={1\over\sqrt{\pi}} \int_{-\infty}^{X/2} e^{-u^2} du  \ \ \ \ \mbox{and }
\ \ \ \ G_2(X,Y) = -{e^{-(X+Y)^2/8}\over2\sqrt{2}\pi} \int_{-\infty}^{X-Y\over
2\sqrt{2}} e^{-u^2} du \,.
\label{scaling-n-bis}
\end{equation}

The scaling form (\ref{scaling-n},\ref{scaling-n-bis}) of the two-point
function  is very reminiscent of  what is known for the  steady state
of the  SSEP on an open interval \cite{Spohn} of length $L$ , where the two point
function scales like  $L^{-1}$ and the $n$-point (connected) correlation
function scales like $L^{1-n}$ \cite{DLS5}.
It would be interesting to know whether the scaling form
(\ref{scaling-n})  remains valid for $n \geq 2$. If  so one could
then try to determine the scaling functions $G_n$ in order to obtain the
large deviation function of an arbitrary density profile
\cite{BDGJL2,DLS2}.

\section{Conclusion}
\label{sec:conc}
In the present work, we have shown that, for a step initial density profile,
the generating function of the integrated current $Q_t$ is a function of a
single parameter $\omega$, defined in (\ref{omega-def}), which takes a simple
closed expression (\ref{res1}).
This $\omega$ dependence is also valid for an arbitrary graph.
In one dimension, for this non-equilibrium initial condition, the distribution
of $Q_t$ is clearly not Gaussian with a tail which decays faster than a
Gaussian (\ref{q3}).

It would be interesting to generalize our results to other diffusive systems
and to try to calculate the distribution of $Q_t$ with the same step initial
condition, in order to see under which conditions one could recover the
non-Gaussian decay (\ref{q3}). The most promising approach, at the moment, is
to try to generalize the macroscopic theory of Bertini, De Sole, Gabrielli,
Jona--Lasinio, and Landim \cite{BDGJL2,BDGJL5,BDGJL6} to a non-steady state
initial condition.

A possible extension of the present work would be to introduce a weak
asymmetry (of order $t^{-1/2}$) in the hopping rates to understand the
cross-over in the current fluctuations between the SSEP and the ASEP.

Another interesting question would be to determine the scaling form
(\ref{scaling-n}) of all the higher correlation functions and to obtain
the large deviation function of an arbitrary density profile, when the
initial condition is, as in figure \ref{initial} far from a steady state
situation.

\appendix

\section{Determinant expression (\ref{sn-bis})  of $\langle e^{\lambda 
Q_t}\rangle$}
\label{Apx:1}
In this appendix we derive the general expression (\ref{sn-bis}) of the
$(s_n)$ from (\ref{sn_1}). Our derivation relies heavily on results obtained
by Tracy and Widom on the particle trajectories of an ASEP model with the same
geometry\cite{TW1,TW2}.
In the present appendix we will use the Bethe ansatz itself (\cite{TW1},
Theorem 2.1),
\begin{equation}\label{TWAnsatz}
  P_t^{(n)}(y_1,y_2,..,y_n|x_1,x_2, .. x_n) =   
  \sum_{\sigma}\mbox{sgn}(\sigma) \oint
  \left[\prod_{k=1}^n {dz_k \over 2i\pi z_k}
  e^{t(z_k+1/z_k-2)}z_{\sigma(k)}^{y_k-x_{\sigma(k)}}\right] \left[\prod_{k<l}
  {z_{\sigma(k)}z_{\sigma(l)}+1- 2z_{\sigma(k)}\over z_k z_l+1-2z_k}\right]\,,
\end{equation}
as well as two algebraic identities (\cite{TW1}, eq. (1.6) and
\cite{TW2}, eq. (7)),
\begin{eqnarray}\label{TWmagic}
  \sum_\sigma \mbox{sgn}(\sigma)\left[\prod_{k=1}^n {z_{\sigma(k)}.. 
  z_{\sigma(n)}\over 1-z_{\sigma(k)}..z_{\sigma(n)}}\right]\left[
  \prod_{k<l}(z_{\sigma(k)}z_{\sigma(l)}+1-2z_{\sigma(k)})\right] &=& z_1..z_n 
  {\prod_{k<l}(z_l-z_k) \over \prod_k (1-z_k)}\\
   \left[\prod_{k=1}^n {1\over (1-z_k)^2}\right]\left[\prod_{k\neq l} 
  {z_k-z_l\over z_k z_l+1-2z_k}\right] &=& \det\left({1\over z_k 
  z_l+1-2z_k}\right)_{1\leq k,l \leq n}\,.\label{TWdet}
\end{eqnarray}

In order to obtain $s_n$ from (\ref{sn_1}), one has to sum
$P_t^{(n)}(y_1,y_2,..,y_n|x_1,x_2, .. x_n)$ over $x_1<..<x_n\leq0$ and $0 <
y_1<..<y_n$ : this yields
$$\sum_{x_1<..<x_n\leq0} \prod_k z_k^{-x_k} = {1\over
z_1..z_n}\left[\prod_{k=1}^n {z_1..z_k \over 1- z_1..z_k} \right] \mbox{ and }
\sum_{0< y_1<..<y_n} \prod_k z_{\sigma(k)}^{y_k} = \prod_{k=1}^n
{z_{\sigma(k)}..z_{\sigma(n)}\over 1-z_{\sigma(k)}..z_{\sigma(n)}} \,.$$

Therefore

$$ s_n = \sum_\sigma \mbox{sgn}(\sigma)\oint \left[\prod_{k=1}^n {dz_k 
 e^{t(z_k+1/z_k-2)} \over 2i\pi z_k^2}{z_1..z_k \over 1-z_1..z_k}
{z_{\sigma(k)}..z_{\sigma(n)} \over 1-z_{\sigma(k)}..z_{\sigma(n)}}\right]
\left[\prod_{k<l} {z_{\sigma(k)}z_{\sigma(l)}+1-2z_{\sigma(k)}
\over z_k z_l+1-2z_k}\right]\,.$$

We are now in position to use (\ref{TWmagic}), which yields

$$ s_n =\oint \left[\prod_{k=1}^n {dz_k e^{t(z_k+1/z_k-2)}
\over 2i\pi z_k(1-z_k)} {z_1..z_k \over 1-z_1..z_k}\right]\left[
\prod_{k<l} {z_l-z_k \over z_k z_l+1-2z_k}\right]\,.$$

In order to eliminate the remaining ${z_1..z_k \over 1-z_1..z_k}$, it is
necessary to use (\ref{TWmagic}) again. As the factors in (\ref{TWmagic}) are
of the form ${z_k..z_n \over 1-z_k..z_n}$, we first relabel $z_k \to z_{n-k}$,
so that

\begin{eqnarray*}
  s_n &=&\oint \left[\prod_{k=1}^n {dz_k e^{t(z_k+1/z_k-2)}
  \over 2i\pi z_k(1-z_k)} {z_k..z_n \over 1-z_k..z_n}
  \right]\left[\prod_{k>l} {z_l-z_k \over z_k z_l+1-2z_k}\right]\\
  &=& \oint \left[\prod_{k=1}^n {dz_k e^{t(z_k+1/z_k-2)}
  \over 2i\pi z_k(1-z_k)} \right]{\prod_{k<l} (z_k-z_l) \over 
  \prod_{k\neq l}(z_k z_l+1-2z_k)}\left[\prod_{k<l}(z_k z_l+1-2z_k)\right] 
  \left[\prod_{k=1}^n{z_k..z_n \over 1-z_k..z_n}\right]\,.
\end{eqnarray*}

We then replace the expression above by its average over all permutations of
the $(z_k)$ in order to apply (\ref{TWmagic}). The first term of the product
is unchanged and the second term gets a sign factor $\mbox{sgn}(\sigma)$.
Hence

\begin{eqnarray*}
  s_n &=& \sum_{\sigma} {\mbox{sgn}(\sigma)\over n!} \oint \left[\prod_{k=1}^n
  {dz_k e^{t(z_k+1/z_k-2)} \over 2i\pi z_k(1-z_k)} \right]\left[\prod_{k<l}
  (z_k-z_l) \right]{\prod_{k<l}(z_{\sigma(k)} z_{\sigma(l)}+1-2z_{\sigma(k)})
  \over \prod_{k\neq l}(z_k z_l+1-2z_k)} \left[\prod_{k=1}^n
  {z_{\sigma(k)}..z_{\sigma(n)} \over 1-z_{\sigma(k)}..z_{\sigma(n)}}\right]\\
  &=& {1 \over n!}\oint \left[\prod_{k=1}^n {dz_k e^{t(z_k+1/z_k-2)} \over
  2i\pi (1-z_k)^2} \right]\left[\prod_{k\neq l} { z_k-z_l \over z_k
  z_l+1-2z_k}\right]\,.
\end{eqnarray*}
Applying (\ref{TWdet}) then leads to
$$s_n(t) ={1\over n!} \oint \left[\prod_{k=1}^n {dz_k\over 2i\pi} \right] \det
\left( {e^{t(z_k+1/z_k-2)}\over z_k z_l+1-2z_k}\right)_{1\leq k,l \leq n}\,,$$
which is the expression (\ref{sn-bis}).

\section{Derivation of the asymptotics (\ref{sn-ter}) of the generating
function }
\label{Apx:2}
In this appendix we derive the large $t$ behavior of the integrals $I_n = {\rm
tr}\,K^n$, which appear in (\ref{logdeterminant},\ref{Kdef},\ref{sn-ter}) .
$I_n$ can be expressed as
$$ I_n = {\rm tr}\,K_t^n= \oint {dz_1\over2i\pi}
\cdots{dz_n\over2i\pi}K_t(z_1,z_2) \cdots K_t(z_n,z_1) = \oint \prod_{k=1}^n
{dz_k\over2i\pi}{e^{t(z_k+1/z_k-2)}\over z_k z_{k+1}+1-2 z_k}\,,$$
where the integration contour of each $z_k$ is a circle of radius $r\ll 1$
(and by convention we have $z_{n+k}\equiv z_k$).
By using the identity 
$$\frac{e^{t(z_{k+1}+1/z_{k}-2)}}{z_{k+1}+1/z_{k}-2} =
\frac{1}{z_{k+1}+1/z_{k}-2} +
\int_0^t dt_k  \ e^{t_k(z_{k+1}+1/z_{k}-2)} \,,$$
$I_n$ can be expressed as
$$I_n = \sum_{E\subset\{1,..,n\}} \oint \left[\prod_{k=1}^n\frac{dz_k}{2i\pi
z_k} \right]\left[ \prod_{k\notin E} \frac{1}{z_{k+1} +1/z_{k}-2} \right]\left[
\prod_{k\in E} \int_0^t dt_k \ e^{t_k(z_{k+1}+1/z_{k}-2)}\right] \,.$$

If $E\neq \{1,..,n\}$, then there exists at last one $k$ such that
$k\not\in E$. For fixed values of the $(z_l)_{l\neq k}$ and the
$(t_l)_{l\neq k}$, the integral over $z_k$ can be written  either as
$$\oint\frac{dz_k}{2i\pi z_k} \ \frac{e^{t_{k-1}z_k}}{z_{k+1}+1/z_k-2} 
\ \ \ \ \ \ \ \ \ \text{or \ as} \ \ \ \  \ \ \ \ \
\oint\frac{dz_k}{2i\pi z_k} \ \frac{1}{(z_{k+1}+1/z_k-2) \
(z_{k}+1/z_{k-1} -2)}\,.
$$
As the integration contour is a  circle of very small radius these
expressions  obviously vanish. 
 Therefore the case $E= \{1,..,n\}$ gives the only non-zero contribution to
$I_n$ :
$$ I_n = \oint_{|z_k|=r}  \int_0^t \prod_{k=1}^n\frac{dz_k  \ dt_k}{2i\pi z_k} 
e^{t_k(z_{k+1}+1/z_{k}-2)} \,.$$
\\
The integrals over the $z_k$'s can now be evaluated using the saddle-point
method around $z_k=\sqrt{t_{k}/t_{k-1}}$. This yields
$$ I_n \simeq \int_0^t \prod_{k=1}^n \frac{dt_k}{2\sqrt{\pi t_k}}
e^{-(\sqrt{t_k}-\sqrt{t_{k-1}})^2} = \left(\frac{t}{\pi}\right)^{n/2}\int_0^1
\prod d\alpha_k e^{-t(\alpha_k-\alpha_{k-1})^2}\,.$$

Finally, since the integrand above is only non-vanishing when
$|\alpha_k-\alpha_l| \sim 1/\sqrt{t}$ for $t\to\infty$, one  gets
$$I_n {\simeq}\left(\frac{t}{\pi}\right)^{n/2}\int_0^1
d\alpha_1 \int_{-\infty}^\infty d\alpha_2 \cdots d\alpha_n
e^{-t\sum(\alpha_k-\alpha_{k-1})^2} = \sqrt{\frac{t}{\pi n}}\,.$$
and this completes the proof of (\ref{sn-ter}).



\end{document}